\documentclass[aps]{revtex4}
\usepackage{graphicx}
\def\bc{\begin{center}}
\def\ec{\end{center}}

\def\beq{\begin{equation}}
\def\eeq{\end{equation}}

\def\d{\downarrow}
\def\u{\uparrow}

\def\br{{\bf r}}

\def\bk{{\bf k}}

\def\bE{{\bf E}}
\def\sgn{{\rm sgn}}

\begin{document}


\title{
Circular edge states in photonic crystals with a Dirac node
}
\author{K. Ziegler}
\affiliation{Institut f\"ur Physik, Universit\"at Augsburg\\
D-86135 Augsburg, Germany}
\date{\today}

\begin{abstract}
Edge states are studied for the two-dimensional Dirac equation in a circular geometry. 
The properties of the two-component electromagnetic field are discussed in terms of the 
three-component polarization field, which can form a vortex structure near the Dirac node
with a vorticity changing with the sign of the Dirac mass. The Berry curvature of the 
polarization field is related to the Berry curvature of the Dirac spinor state. This 
quantity is sensitive to a change of boundary conditions. In particular, it vanishes 
for a geometry with a single boundary but not for a geometry with two boundaries.
This effect is robust against the creation of a step-like edge inside the sample.
\end{abstract}

\maketitle

\section{Introduction}

Recent progress in the design of photonic metamaterials has opened a new research area
in which many ideas, originally developed for quantum field theories or for 
electronic systems, can be realized and tested with photons under much more general
conditions than originally anticipated. A crucial advantage of photonics in comparison to
electronics is that photons are not charged and do not interact directly with each other. 
Moreover, photons cover a wide range
of length scales which are described by the same Maxwell theory. This provides the 
opportunity to design metamaterials from nano-- to centimeter length scales for the same
type of physics. For example, there have been experiments with visible light 
as well as with microwaves in the frequency regime down to 10 GHz.
Thus, the characteristic lengths in the photonic experiments vary
from $10^{-9}$m ... $10^{-2}$m. 

The analogy between electronic materials and photonic systems in dielectric media has lead to
a number of interesting experiments. Starting from the popular case of graphene, where the 
underlying honeycomb lattice is formed by carbon atoms, exceptionally robust electronic 
transport properties have been observed \cite{novoselov05}. 
This is caused by spectral degeneracies in the form of Dirac nodes, which were also 
studied recently in photonic crystals \cite{haldane08,zhang08,ochiai09,wang09,zandbergen10,
huang11,rechtsman13,ma15,titum15,cheng16}. In analogy to the robust electronic 
transport, the photonic experiments indicate very robust photonic edge states, protected by
topological invariants, such as the Chern number \cite{haldane08}.  
However, in the presence of time-reversal symmetry, as it exists in photonic lattices with a real dielectric
constant, Dirac nodes appear only pairwise with opposite sign of the Dirac mass. This implies that the
Chern number, which is an integral over the entire Brillouin zone, vanishes in this case.
Breaking this symmetry, for instance through the Faraday effect, can result in a change of 
the Chern number \cite{haldane08}. Several procedures are
discussed in the recent literature, how this can be achieved in photonic crystals and metamaterials
\cite{huang11,rechtsman13,titum15,ma15,cheng16}. A typical procedure for the creation of 
edge states in systems with Dirac nodes consists of three steps: (i) opening of a gap at the Dirac nodes,
(ii) changing the Chern number by breaking the time-reversal symmetry and (iii) removing all Dirac
nodes except for a single one. Although the last step is not necessary, it is possible
\cite{haldane88,campana06,hill11} and simplifies the theoretical discussion.
 
The following study has been motivated by a series of experiments with edge states in photonic crystals near
a slightly gapped Dirac node \cite{huang11,ma15,cheng16}. Depending on the sign of the gap parameter (or Dirac mass),
there is a positive or a negative Chern number associated with this sign \cite{haldane08}. The Dirac mass
may change in space such that positive and negative spatial regions can be distinguished. 
It is possible to design a 
photonic crystal with several regions of positive and negative Dirac mass, and edge states between these regions
\cite{cheng16}. The aim of this work is to study the properties of edge states in different geometries,
how they are affected by the geometry of the boundaries,
and to propose a method how to characterize them through the polarization of the electromagnetic field. 

In the following we distinguish two different types of edges, namely sample boundaries and edges created by
a sign change of the Dirac mass. For clarity, we will use the name edge only for the latter but reserve
boundaries for the sample geometry. In agreement with the experimental studies,
we consider a two-dimensional photonic crystal with a single Dirac node 
\cite{haldane08,zhang08,ochiai09,wang09,zandbergen10,huang11,rechtsman13,ma15,titum15,cheng16} 
and assume a small gap.
The two-component electromagnetic field in the $x$--$y$ plane is a solution of the 
two-dimensional (2D) Dirac equation and represents a Dirac spinor in a formal sense.  
The solutions of the corresponding Helmholtz equation are well known and can be found in text 
books (e.g., Ref. \cite{courant}). 
Although the purpose of this short note is to describe edge states in a photonic
crystal near a Dirac node, the results should also be applicable to electronic
quasiparticles with a cone-like spectrum. However, the effect of the Coulomb interaction must
be included in the electronic case, which will not be considered here.  

The Berry phase \cite{berry84} or the Chern number, both obtained by integration,
have been used to characterize the properties of the spinor eigenstate of the Hamiltonian as a 
function of the Dirac mass and the band index (upper or lower band) \cite{haldane08,degail12,cheng16}.
Here we employ an alternative approach with the polarization field. This is a three-component 
rather than a two-component vector. A connection between the polarization and the Berry
curvature was discussed earlier for an optical fiber in Ref. \cite{haldane86}.

\section{2D Dirac equation}

We consider a photonic crystal which is characterized by a periodic space-dependent 
dielectric tensor $\epsilon$. This implies that the spectrum of the Maxwellian 
has a band structure \cite{john87,ph_lattice}. As mentioned in the Introduction, it was shown that
for a hexagonal arrangement of dielectric rods or cylinders 
there are two spectral nodes {\bf at $\omega=\omega_D$} in the band structure \cite{haldane08,huang11}. 
Near these spectral nodes the Maxwell equation 
is reduced for the TE (transverse electric) state to two 2D Dirac equations
\beq
H_{D;\pm}{\bf E}_\pm=0 \ ,
\ \ \ 
H_{D;\pm}=\pmatrix{
m & i\partial_x\pm\partial_y \cr
i\partial_x\mp\partial_y & -m \cr
}
\label{dirac00}
\eeq
$\partial_x\equiv \partial/\partial x$, ...\ . All lengths are measured in units
of $c/\sqrt{\epsilon}\omega$, where $c$ is the speed of light, $\omega$ the frequency
of the electromagnetic field and $\epsilon$ the dielectric constant.
The degeneracy of the two Dirac equations with $H_{D;\pm}$ can be lifted by 
introducing different masses $m_\pm$, where one mass is much bigger than the other. 
For a proper frequency 
$\omega_D$ there is only one Dirac node (with massless or almost massless) Dirac 
particles \cite{haldane08,huang11}. The eigenvalue problem of the Dirac operator
\beq
H_D{\bf E}=
\pmatrix{ 
m & i\partial_x+\partial_y \cr
i\partial_x-\partial_y & -m \cr
}\pmatrix{
E_\u \cr
E_\d \cr
}=E\pmatrix{
E_\u \cr
E_\d \cr
}
\label{dirac00}
\eeq
on a square geometry is solved by
\beq
\bE=\pmatrix{
E_\u \cr
E_\d \cr
}
=\frac{1}{\sqrt{2}}
\pmatrix{
\frac{k_x-ik_y}{-E+m} \cr
1 \cr
}e^{ik_x x+ik_y y}
\label{plane_wave}
\eeq
with the condition
\beq
E^2=k_x^2+k_y^2+m^2
\eeq
for the wavevector ${\bf k}=(k_x,k_y)$, which is adapted to the boundary conditions.
This result describes two-dimensional Dirac photons for energies $|E|\ge |m|$.

An electromagnetic field is characterized by four Stokes parameters \cite{hulst81,mishchenko06}.
They can be expressed as quadratic forms of the electric field $\bE$ with Pauli
matrices $\sigma_j$ ($j=0,x,y,z$; $\sigma_0$ is the $2\times 2$ unit matrix):
$I=(\bE\cdot\sigma_0\bE)$ is the intensity with the two-dimensional scalar product $(.\cdot .)$
and 
\beq
Q=(\bE\cdot\sigma_z\bE)
\ , \ \ \ 
U=(\bE\cdot\sigma_x\bE)
\ , \ \ \ 
V=(\bE\cdot\sigma_y\bE)
\label{stokes00}
\eeq
are the other Stokes parameters, which provide the polarization. 
It should be noticed that the relation $Q^2+U^2+V^2=I^2$ implies that the tip of
the vector $(Q,U,V)$ describes a sphere of radius $I$ (Poincar\'e sphere).
After normalizing the radius to 1, the vector field
\beq
{\bf n}=\frac{1}{I}\pmatrix{
U \cr
V \cr
Q \cr
}
\label{norm_pol}
\eeq
can be used to characterize the polarization through a Berry curvature.
This will be briefly discussed in Sect. \ref{sect:discussion}.  

For the plane wave solution of Eq. (\ref{plane_wave}) we get for the Stokes
parameters in $\bk$--space expressions which are uniform in $\br$--space
\beq
I=\frac{E}{E-m}
\ ,\ \ \ 
\pmatrix{
U\cr
V\cr
}=-\frac{\bk}{E-m}
\ ,\ \ \ 
Q=\frac{m}{E-m}
\eeq
as the result of the translational invariance of the system and the square geometry.
In the presence of an edge or in a circular geometry this translational invariance is broken.
As a consequence, the Stokes parameters will become $\br$ dependent. 
\begin{figure}[t]
\begin{center}
\includegraphics[width=7cm,height=3.3cm]{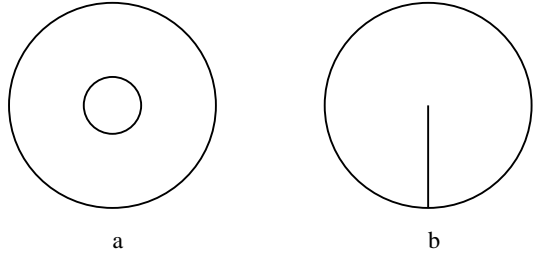}
\caption{\label{fig:samples}
Geometry of circular samples. Sample a has two circular boundaries, whereas sample b
with the radial cut has a single boundary.
}
\end{center}
\end{figure}

\subsection{Circular geometry}

First, we consider states outside the gap (i.e., for $E^2>m^2$) in a disc geometry.
The Dirac eq. (\ref{dirac00}) reads in terms of polar coordinates $(x,y)=(r\cos\alpha,r\sin\alpha)$
\beq
\pmatrix{
m & ie^{-i\alpha}(\partial_r-\frac{i}{r}\partial_\alpha) \cr
ie^{i\alpha}(\partial_r+\frac{i}{r}\partial_\alpha) & -m \cr
}\pmatrix{
E_1 \cr
E_2 \cr
}=E\pmatrix{
E_1 \cr
E_2 \cr
}
\ .
\label{dirac01}
\eeq
Assuming that the solution $E_1$ is rotational invariant
at the boundary of the disc, this equation is solved by
\beq
\pmatrix{
E_1 \cr
E_2 \cr
}=\pmatrix{
a J_0({\bar r})+b Y_0({\bar r}) \cr
ie^{i\alpha}\frac{\sqrt{E^2-m^2}}{E-m} [a J_1({\bar r})+ b Y_1({\bar r})] \cr
}
\ ,
\eeq
where ${\bar r}=\sqrt{E^2-m^2}r$, and $J_n$, $Y_n$ are Bessel functions. 
The coefficients $a$ and $b$ must be fixed by the specific boundary conditions
of $\bE$. Then the Stokes parameters of this solution are
\beq
I=(aJ_0+bY_1)^2+(aJ_1+bY_1)^2
\ , \ \ \ 
\pmatrix{
U \cr
V \cr
}
=2\frac{\sqrt{E^2-m^2}}{r(E-m)}(aJ_0+bY_1)(aJ_1+bY_1)
\pmatrix{
-y \cr
x \cr
}
\eeq
and
\beq
Q=(aJ_0+bY_1)^2-(aJ_1+bY_1)^2
\ .
\eeq
$I$ and $Q$ are rotational invariant (i.e., they do not depend on $\alpha$),
and the other two Stokes parameter describe a circulating polarization
in the $x$--$y$ plane around the center of the disc. Since the sign of 
the vorticity changes as a function of the radius because of changing
signs of the Bessel functions, this structure does not 
represent a vortex.

The situation is different inside the gap (i.e., for $E^2<m^2$), where states 
decay exponentially. For example, we consider the zero energy state of the Dirac 
eq. (\ref{dirac01}) for the two geometries of Fig. \ref{fig:samples}.
For Fig. \ref{fig:samples}a with two boundaries we assume
$E_1$ is rotational invariant at both boundaries. 
Then the solution reads 
\beq
\pmatrix{
E_1 \cr
E_2 \cr
}=\pmatrix{
K_0({\bar r}) \cr
-ie^{i\alpha}\sgn  m\ K_1({\bar r}) \cr
}
\ ,
\eeq
where $K_n$ are modified Bessel functions and ${\bar r}=|m|r$.
The inner circle in Fig. \ref{fig:samples}a is necessary, since $K_0$, $K_1$ diverge
for $r\sim0$. In more physical terms, the inner boundary creates the edge state. 
The Stokes parameters read in this case
\beq
I=K_0^2+K_1^2
\ , \ \ \ 
\pmatrix{
U \cr
V \cr
}
=-2\sgn  m\ \frac{K_0K_1}{r}\pmatrix{
-y \cr
x \cr
}
\ ,\ \ \ 
Q=K_0^2-K_1^2
\ ,
\label{stokes01}
\eeq
where $U$ and $V$ represent a vortex solution in the plane
with a vorticity proportional to $\sgn m$, since $K_0K_1>0$.  

For the simply connected geometry with a radial cut of Fig. \ref{fig:samples}b 
the Dirac eq. (\ref{dirac01}) is solved by
\beq
\pmatrix{
E_1 \cr
E_2 \cr
}=\frac{1}{\sqrt{2}}\pmatrix{
e^{-i\alpha/2} \cr
-i\sgn  m\ e^{i\alpha/2} \cr
}\frac{e^{-|m| r}}{\sqrt{r}}
\ ,
\label{sol_scg}
\eeq
where the angular function $e^{\pm i\alpha/2}=\sqrt{x\pm i y}/\sqrt{r}$ has a branch cut. This must
be associated with boundary conditions such that the phase of the electric field has a $\pi$ jump at
the radial cut in Fig \ref{fig:samples}b. The intensity and the other Stokes parameters are not 
affected by the branch cut: 
\beq
I=\frac{e^{-2|m| r}}{r}
\ ,\ \ \ 
\pmatrix{
U \cr
V \cr
}
=-\sgn  m\ \pmatrix{
-y \cr
x \cr
}\frac{e^{-2|m|r}}{r^2}
\ , \ \ \ 
Q=0
\ .
\label{stokes02}
\eeq
Again, $U$, $V$ represent a vortex in the plane with a vorticity proportional
to the sign of the Dirac mass. Thus, the Stokes parameters agree qualitatively for the 
two circular geometries. However, the parameter $Q$ (the third component of 
the polarization, according to Eq. (\ref{stokes00})) vanishes only for the simply
connected geometry, whereas $Q<0$ for the geometry with two boundaries.

\section{Edge states with 2D Dirac photons}
 
\subsection{Straight edge}

\begin{figure}
\begin{center}
\includegraphics[width=7cm,height=2.5cm]{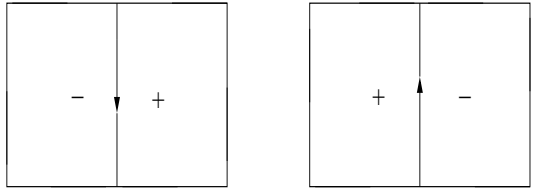}
\caption{\label{fig:straight_edge}
A straight edge, where the arrows indicate the
polarization. The $\pm$ are the signs of the Dirac mass.
}
\end{center}
\end{figure}

Now we consider a space-dependent Dirac mass which forms an edge along the $y$-direction with
$-m$ ($m$) for $x<0$ ($x\ge0$), as visualized in Fig. \ref{fig:straight_edge}.
This straight edge creates a zero energy state \cite{ziegler17}
\beq
\pmatrix{
E_1 \cr
E_2 \cr
}=\frac{1}{\sqrt{2}}\pmatrix{
i\sgn \ m \cr
1\cr
}e^{-|mx|}
\eeq
inside the gap of width $2|m|$. For this edge state we have $Q=0$,
and the remaining Stokes parameters describe a global polarization in the $x$--$y$ plane:
\beq
I=e^{-2|mx|}
\ ,\ \ \ 
\pmatrix{
U\cr
V\cr
}
=-\pmatrix{
0 \cr
\sgn \ m \cr
}e^{-2|mx|}
\ , \ \ \
Q=0
\ .
\label{stokes_straight}
\eeq
Thus, the polarization is directed along the $y$--axis, where the sign of $m$ 
determines its orientation (cf. Fig. \ref{fig:straight_edge}).

\subsection{Circular edge for a geometry with two circular boundaries}
\label{sect:2_boundaries}

\begin{figure}[t]
\begin{center}
\includegraphics[width=7cm,height=3.3cm]{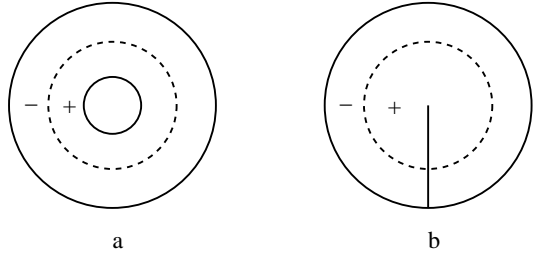}
\caption{\label{fig:samples_2}
Geometry of circular samples of Fig. \ref{fig:samples} with a circular edge (dashed circle).
}
\end{center}
\end{figure}

Now we return to the circular geometries of Fig. \ref{fig:samples} and add
an edge that forms a loop inside a sample. 
For simplicity, we assume a concentric circular edge of radius $r_0$, where the Dirac 
mass $m$ jumps from the inside ($m=m_i$) to the outside region ($m=-m_i$),
as sketched by the dashed circles in Fig. \ref{fig:samples_2}. Then the 
zero energy state of the Dirac eq. (\ref{dirac01}) is solved
by the spinor $\bE$ with components
\beq
\pmatrix{
E_1 \cr
E_2 \cr
}=\pmatrix{
aK_0({\bar r})+bI_0({\bar r}) \cr
ie^{i\alpha}\sgn  m_i[-aK_1({\bar r})+bI_1({\bar r})] \cr
}
\ \ \ \ 
(r\le r_0)
\label{sol1}
\eeq 
and
\beq
\pmatrix{
E_1 \cr
E_2 \cr
}=\pmatrix{
cK_0({\bar r}) \cr
ie^{i\alpha}\sgn  m_i cK_1({\bar r}) \cr
}
\ \ \ \ 
(r>r_0)
\ .
\label{sol2}
\eeq
$K_n$ and $I_n$ are modified Bessel functions, ${\bar r}=|m_i|r$. The coefficients
\beq
a=\frac{I_1({\bar r}_0)K_0({\bar r}_0)-I_0({\bar r}_0)K_1({\bar r}_0)}
{I_1({\bar r}_0)K_0({\bar r}_0)+I_0({\bar r}_0)K_1({\bar r}_0)}
\ ,\ \ \ 
b=\frac{2c}{I_0({\bar r}_0)/K_0({\bar r}_0)+I_1({\bar r}_0)/K_1({\bar r}_0)}
\eeq
secure a continuous behavior of the spinor at the edge $r=r_0$, and the coefficient
$c$ must be fixed by the boundary value of $\bE$ at the inner circular boundary.

\begin{figure}[t]
\begin{center}
\includegraphics[width=9cm,height=6cm]{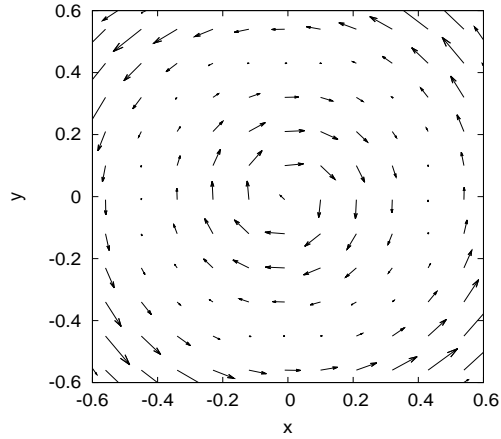}
\caption{\label{fig:polarization_a}
The $U$--$V$ projection of the polarization field for positive $m_i$ inside the circular edge
with radius $r_0=1$ for the geometry of Fig. \ref{fig:samples_2}a: the edge, 
together with the inner circular boundary, creates a double vortex.
(Sect. \ref{sect:2_boundaries}). The polarization is normalized by the intensity here.
}
\end{center}
\end{figure}

From the solutions (\ref{sol1}, (\ref{sol2}) we obtain the Stokes parameters as
\beq
I=\cases{
(aK_0+bI_0)^2+(-aK_1+bI_1)^2 & for $r\le r_0$ \cr
c^2(K_0^2+K_1^2) & for $r > r_0$ \cr \cr
}
\ ,
\label{intensity1}
\eeq
\beq
Q=\cases{
(aK_0+bI_0)^2-(-aK_1+bI_1)^2  & for $r\le r_0$ \cr
c^2(K_0^2-K_1^2) & for $r > r_0$ \cr
}
\ ,
\eeq
and for the polarization in the $U$--$V$ plane we get
\beq
\pmatrix{
U \cr
V \cr
}
=P({\bar r})\pmatrix{
-y \cr
x \cr
}
\ , \ \ \ 
P =\frac{2\sgn m_i}{r}\cases{
(aK_0+bI_0)(-aK_1+bI_1) & for $r\le r_0$ \cr \cr
c^2K_0K_1 & for $r > r_0$ \cr \cr
}
\ .
\label{polar3}
\eeq
This agrees for $r>r_0$ with the result for circular system without an edge in Eq. (\ref{stokes01}),
since the Dirac mass is $-m_i$.
Inside the edge for $r\le r_0$ the situation is different, though. As visualized in Fig. \ref{fig:polarization_a},
the vorticity changes from a counterclockwise vorticity to a clockwise vorticity as
we get closer to the center of the circle, because the Dirac mass is $m_i>0$ and the factor
$(aK_0+bI_0)(-aK_1+bI_1)$ becomes negative for small ${\bar r}$. Near the circular edge, however, there
is a counterclockwise vorticity. Therefore, the vorticity near the edge behaves in analogy with
the orientation of the polarization near a straight edge (cf. Eq. (\ref{stokes_straight}) and Fig.
\ref{fig:straight_edge}).


\subsection{Simply connected geometry with a circular edge}
\label{sect:1_boundary}

\begin{figure}[t]
\begin{center}
\includegraphics[width=9cm,height=6cm]{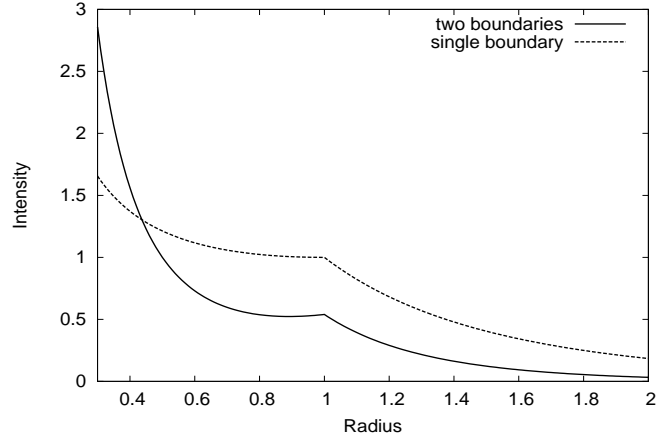}
\caption{\label{fig:intensity}
The radial intensity profile has a cusp from the edge at $r_0=1$, according to the
expressions (\ref{intensity1}) and (\ref{intensity2}). 
}
\end{center}
\end{figure}

Adding a concentric circular edge to the geometry of 
Fig. \ref{fig:samples}b results again in two regions which are separated by the edge 
(cf. Fig. \ref{fig:samples_2}b).
In order to get a continuous spinor $\bE$ from the solution (\ref{sol_scg}),
the mass sign of the second component must be compensated.
Moreover, an extra factor must take care of the fact that the exponential
function is continuous at the edge, which is possible by replacing
\beq
e^{-|m_i|r}\to e^{-|m_i||r-r_0|}
\ .
\eeq
Thus, we obtain
\beq
\pmatrix{
E_1 \cr
E_2 \cr
}=\frac{1}{\sqrt{2}}\pmatrix{
e^{-i\alpha/2} \cr
i\sgn m_i\ e^{i\alpha/2} \cr
}\frac{e^{-|m_i| |r-r_0|}}{\sqrt{r}}
\label{sol_scg_2}
\eeq
as the zero energy solution of the Dirac equation.
Despite of the radial cut,
the intensity has no angular dependence and reads
\beq
I=\frac{e^{-2|m_i||r-r_0|}}{r}
\ .
\label{intensity2}
\eeq
As a typical example the radial intensity profile is plotted in Fig. \ref{fig:intensity} 
with a cusp at the edge. The intensity decays exponentially on the scale $1/|m_i|$ away 
from the edge. On the other hand, at the center $r=0$ the intensity diverges like $1/r$,
although its spatial integral (i.e., the total intensity) is finite:
\beq
\int_0^\infty I rdr=\frac{2-e^{-2|m_i|r_0}}{|m_i|}
\ .
\eeq
Moreover, $Q=0$ and the remaining two Stokes parameters read
\beq
\pmatrix{
U \cr
V \cr
}
=\sgn  m_i\pmatrix{
-y \cr
x \cr
}\frac{e^{-2|m_i||r-r_0|}}{r^2}
\ ,
\eeq
representing a vortex whose vorticity changes with the sign of the Dirac mass. 
In contrast to the case with two boundaries of Sect. \ref{sect:2_boundaries}, 
the vorticity does not change its sign as a function of the radius.
Moreover, the Stokes parameters are not affected by the radial cut, which is
similar to the result for the Dirac eq. without an edge in Eq. (\ref{stokes02}).

\section{Discussion}
\label{sect:discussion}

The polarization of a 2D two-component electromagnetic field near a single gapped Dirac cone
is analyzed under different boundary conditions, with and without a circular edge. 
Without an edge there are two characteristic solutions, depending on the sample
geometry proposed in Fig. \ref{fig:samples}: For a sample with a hole there is
a three-component polarization field $(U,V,Q)$ with a vortex in the $U$--$V$ plane. The solution for
the geometry with a single boundary, on the other hand, is a two-component polarization vortex field 
inside the $U$--$V$ plane. The vorticity depends in both cases on the
sign of the Dirac mass. This is important, since neither the Dirac spectrum, the 
intensity $|\bE|^2$ nor the Stokes parameter $Q$ depend on the sign of the Dirac 
mass. Therefore, the polarization inside the $U$--$V$ plane plays a similar role as 
the Berry connection or the Chern number for the spinor field $\bE$ as a
characterization of the mass sign.

An edge in the form of a step-like sign change of the Dirac mass is usually accompanied
by the formation of an edge state, which decays exponentially off away from the edge.
This is also observed in the case of a circular geometry. In our examples there are
several edge effects:
The vortex is only affected by the edge in the case of the geometry with two boundaries,
as visualized in Fig. \ref{fig:polarization_a}, whereas for the geometry with a single
boundary the vortex structure is the same with or without the edge. The reason
for this difference is that in the former case there are three competing effective edges,
namely the two boundaries and the edge of the changing mass. This leads to the formation of
two different vorticities near the edge. 

The structure of the polarization can also be characterized by the Berry curvature
\cite{berry84,haldane86}
\beq
B=\frac{1}{2}{\bf n}\cdot \left[ \partial_x {\bf n}\times \partial_y{\bf n}\right]
\ ,
\eeq
where ${\bf n}$ is the normalized polarization on the unit sphere of Eq. (\ref{norm_pol}).
$B$ vanishes for the simply connected geometry with and without an edge because of $Q=0$,
meaning that the polarization is flat.
On the other hand, $B$ is a non-vanishing function of the radius $r$ for the geometry with 
two boundaries.
The Berry connection of the spinor field $\bE$ is non-zero in both
cases, though, and it corresponds with the projection of the polarization vortex onto 
the $U$--$V$ plane.

As mentioned in the Introduction, the results of these calculations are applicable to
electronic systems with a Dirac node when the Coulomb interaction is neglected. Then the
Stokes parameters $(U,V,Q)$ are local current densities defined in Eq. (\ref{stokes00})
with $\bE$ replaced by the electronic spinor, since the components of the current 
operator are proportional to the Pauli matrices.
Another interesting realization of circular edges are systems with polaritons \cite{karzig15}.

\section{Conclusions}

Starting from the Dirac eq. (\ref{dirac00}) for a two-component electromagnetic field $\bE$,
we have discussed solutions in the presence of two different types of boundaries 
(cf. Fig. \ref{fig:samples}) and with a circular edge caused by a mass sign change 
(cf. Fig. \ref{fig:samples_2}). From the electromagnetic field $\bE$ the three-component
polarization field was calculated with the result that a boundary creates a 
polarization field with a vortex, whose vorticity is proportional to the sign
of the Dirac mass. In the presence of two boundaries and an edge the competition of
regions with different signs of the Dirac mass creates a structure with changing
vorticities (cf. Fig. \ref{fig:polarization_a}). The Berry curvature of the
polarization field depends on whether there is a sample geometry with one or two
boundaries. This indicates that the polarization field carries substantial information 
which can be a used for measuring the properties of the edge states as well as for
storing information by controlling the boundaries.

\vskip0.3cm

Acknowledgment: I am grateful to A. Genack for introducing me to his experiments
on microwave metamaterials.
This work was supported by a grant of the Julian Schwinger Foundation.

\end{document}